\documentclass[epj]{webofc2}

\pdfoutput=1

\usepackage[varg]{txfonts}   
\usepackage{bbm}
\usepackage{color}

\wocname{EPJ Web of Conferences}
\woctitle{ICNFP 2016}
\newcommand{\EB}{E_\mathrm{B}}
\newcommand{\EE}{E_{\theta}}
\newcommand{\vw}{v_{\mathrm{w}}}
\newcommand{\SU}{\mathrm{SU}}
\newcommand{\SO}{\mathrm{SO}}
\newcommand{\Sp}{\mathrm{Sp}}
\newcommand{\SUL}{\mathrm{SU}(2)_{\mathrm{L}}}
\newcommand{\SUR}{\mathrm{SU}(2)_{\mathrm{R}}}
\newcommand{\ii}{\mathrm{i}}

\newcommand{\XE}[3]{{#1}^{#2}\cdot {#3}}

\begin{document}
\selectlanguage{english}
\title{Vacuum alignment and radiatively induced Fermi scale}

\author{Tommi Alanne\inst{1}\fnsep\thanks{\email{alanne@cp3.sdu.dk}} 
}

\institute{CP$^{3}$-Origins, University of Southern Denmark, Campusvej 55, DK-5230 Odense M, Denmark }

\abstract{%
We extend the discussion about vacuum misalignment by quantum corrections in models with composite pseudo-Goldstone Higgs boson to renormalisable models with elementary scalars. 
As a concrete example, we propose a framework, where the hierarchy between the unification and the Fermi scale emerges radiatively.
This scenario provides an interesting link between the unification and Fermi scale physics.

}
\maketitle
%
\section{Introduction}
\label{sec:intro}

In models with enhanced global symmetries, radiative effects have to be taken into account to determine the actual alignment between the electroweak (EW) gauge group and the stability group related to the spontaneous global symmetry breaking. This was first observed in the context of technicolour by Peskin~\cite{Peskin:1980gc} and Preskill~\cite{Preskill:1980mz}: They discovered that the EW gauge group prefers to be unbroken, and thus tends to destabilise the technicolour vacuum. Later, it was noticed that the corrections from the Standard-Model (SM) fermion sector, most notably the top quark, prefer the fully broken EW gauge group, see e.g.~\cite{Galloway:2010bp,Cacciapaglia:2014uja}, and therefore the actual vacuum alignment depends strongly on the underlying model. Kaplan and Georgi  \cite{Kaplan:1983fs,Kaplan:1983sm} realised that the vacuum misalignment problem in technicolour models can be used to realise the Higgs doublet of the SM  as doublet of dynamically generated Goldstone bosons (GBs). Furthermore, since in such scenarios the Fermi scale, $\vw=246$ GeV, originates due to the vacuum misalignment, the actual compositeness scale can be much higher than in traditional technicolour models.

This phenomenon is not only relevant for underlying composite dynamics, but is present also in models with elementary scalars whenever the scalar potential carries an enhaced global symmetry. It turns out, however, that assuming renormalisability changes the vacuum alignment phenomenology with respect to composite-Higgs case~\cite{Alanne:2016mmn}. Analysing the vacuum alignment issue in the framework of renormalisable models, thus, opens up interesting prospects for models with an elementary pseudo-GB (pGB) Higgs boson and radiative Fermi scale. 

The discussion is organised as follows: In Sec.~\ref{sec:vacAlign}, we outline the vacuum alignment analysis in the framework of renormalisable models with elementary scalars and make a comparison with the composite case. In Sec.~\ref{sec:unification}, we provide in the Pati--Salam-unification framework a concrete model, where the Fermi scale is radiatively generated and has an origin near the unification scale. Finally, we conclude in Sec.~\ref{sec:conc}.

\section{Vacuum alignment with elementary scalars}
\label{sec:vacAlign}

    Whenever the scalar sector features a larger global symmetry than the full Lagrangian, radiative effects must be taken into account to find out the actual alignment of the vacuum with respect to the gauge group. It is well known that in technicolour models, the EW gauge group prefers to be unbroken~\cite{Peskin:1980gc,Preskill:1980mz}, but the actual vacuum alignment crucially depends on the embedding of the SM-fermion masses. 
    
In~\cite{Alanne:2016mmn} it was shown that the different UV structures of the composite-Higgs models and renormalisable scenarios with elementary scalars lead to different vacuum alignment phenomenology. This can be understood in terms of different effective-potential structures: In the composite case, the low-energy effective theory below the compositeness scale depends on the underlying strongly-interacting model and the SM-fermion-mass mechanism, and the leading contributions to the effective potential are quadratic in the compositeness scale, i.e.
\begin{equation}
    \label{eq:}
    V^{\mathrm{eff}}_{\mathrm{comp}}\sim M^2\Lambda^2+\dots
\end{equation}
On the other hand, in a renormalisable model with elementary scalars, the radiative corrections are calculable, and the one-loop
scalar potential is of the form of the renormalised Coleman--Weinberg potential, i.e.
\begin{equation}
    \label{eq:}
    V^{\mathrm{1-loop}}_{\mathrm{elem}}\sim M^4(\log\frac{M^2}{\mu_0^2}-C).
\end{equation}

In either case, the true vacuum depends on the underlying model, and should be determined when the radiative corrections are correctly evaluated. In general, the actual vacuum is a linear combination of the vacuum preserving the EW gauge group, $E_0$, and the one fully breaking it to electromagnetism, $E_{\mathrm{B}}$. In the former case, the stability group related to the global symmetry breaking fully contains the EW gauge group, whereas in the latter case the two subgroups are maximally misaligned. It is convenient to parameterise this misalignment by an angle, $\theta$, and thus write the true vacuum as
\begin{equation}
    \label{eq:vacE}
    \EE=\cos\theta E_0+\sin\theta\EB.
\end{equation}
The value of the angle, $\theta$, is undetermined at the tree-level, but a preferred value is picked once radiative corrections are taken into account. 
The amount of the breaking of the EW subgroup in the spontaneous symmetry breaking thus depends on the vacuum  alignment, and the Fermi scale is given by $\vw=v \sin\theta$. In particular, if a non-zero but small value of the angle, $\theta\ll 1$,  is preferred, the Fermi scale is much smaller than the actual symmetry breaking scale, $v$.

\subsection{$\mathrm{SO}(N)\rightarrow \mathrm{SO}(N-1)$ template}
\label{sec:SON}

To illustrate the main features of the vacuum alignment problem with elementary scalars, let us consider the $\SO(N)\rightarrow\SO(N-1)$ breaking pattern. The $\SO(N)$-symmetric scalar sector can be written as
\begin{equation}
    V_0  = \frac{m^2}{2} \Phi^{\dagger}\Phi + \frac{\lambda}{4!}( \Phi^{\dagger}\Phi)^2.
\end{equation}    
When the scalar $\Phi$ acquires a vev, the symmetry is broken spontaneously to $\SO(N-1)$ leaving behind $N-1$ GBs. It is convenient to parameterise $\Phi$ in terms of these GBs and an $\SO(N-1)$ singlet, $\sigma$, as 
\begin{equation}
    \label{eq:Phi}
    \Phi=\left(\sigma+\ii\Pi^aX^a\right)E,
\end{equation}
where $X^a$ are the broken generators corresponding to the vacuum alingment, $E$.

Whenever $N>4$, we can embed $\SO(4)\cong\SU(2)_{\mathrm{L}}\times\SU(2)_{\mathrm{R}}$ as a subgroup of $\SO(N-1)$ so that EW is unbroken at the tree level. We gauge the EW subgroup by introducing the covariant derivative
	\begin{equation}
	    \label{eq:}
	    D_{\mu}\Phi=\partial_{\mu}\Phi-\ii g \,A_{\mu}^a\,\XE{\tau}{a}{\Phi},
	\end{equation}
	and parameterise the true vacuum as in Eq.~\eqref{eq:vacE}, $ E\equiv \EE=\cos\theta E_0+\sin\theta\EB$, where $E_0$ preserves EW 
	and $\EB$ breaks it to $\mathrm{U}(1)_{Q}$.
	Thus, when $\Phi$ acquires a vev, $\langle\Phi\rangle=v E$, the gauge bosons obtain masses 
	\begin{equation}
		\mu^2_{W}=\frac{1}{4}g^2v^2\sin^2\theta,\quad 
		\mu^2_{Z}=\frac{1}{4}(g^2+g^{\prime\, 2})v^2\sin^2\theta. 
	\end{equation}

Furthermore, we add the SM-like Yukawa interaction for the top quark, i.e. couple it to the left doublet of $\Phi$, such that as $\langle\Phi\rangle=v E$, the top quark gets a mass
\begin{equation}
    \label{eq:}
    m_t=\frac{1}{\sqrt{2}}y_t v\sin\theta.
\end{equation}

To study the vacuum structure, we calculate the one-loop Coleman--Weinberg potential in the $\overline{\mathrm{MS}}$ scheme,
	\begin{equation}
	    \label{eq:Veff}
	    V_{\mathrm{1-loop}}=V_1^{\mathrm{scalar}}+V_1^{\mathrm{gauge}}+V_1^{\mathrm{fermion}},
	\end{equation}
	where
	\begin{align}
	    \label{eq:}
	    V_{1}^{\mathrm{scalar}}=&\frac{1}{64\pi^2}\mathrm{Tr}\left[M^4(\varphi)\left(\log\frac{M^2(\varphi)}{\mu_0^2}-\frac{3}{2}\right)\right],\nonumber\\
	    V_{1}^{\mathrm{gauge}}=&\frac{3}{64\pi^2}\mathrm{Tr}\left[\mu^4(\varphi)\left(\log\frac{\mu^2(\varphi)}{\mu_0^2}-\frac{5}{6}\right)\right],\\
		V_{1}^{\mathrm{fermion}}=&-\frac{4}{64\pi^2}\mathrm{Tr}\left[\left(m^{\dagger}(\varphi)m(\varphi)\right)^2
		    \left(\log\frac{m^{\dagger}(\varphi)m(\varphi)}{\mu_0^2}-\frac{3}{2}\right)\right],
	    \nonumber
	\end{align}
	and $M(\varphi)$, $\mu(\varphi)$, and $m(\varphi)$ are the background-dependent scalar, gauge boson, and fermion mass matrices, 
	respectively.

	To find out the preferred value for the vacuum angle, $\theta$, at the vacuum, we minimize the  the full effective potential, 
	$V_{\mathrm{eff}}=V_0+V_{\mathrm{1-loop}}$ both with respect to the fields and the angle. 
As described in detail in~\cite{Alanne:2016mmn}, the minimum of the potential is at $\theta=0$. Therefore, the EW symmetry remains intact, and there are no non-trivial solutions that would allow for a pGB Higgs.

    \subsection{Vacuum misalignment}
    \label{sec:singl}
In order to break the EW symmetry and obtain a pGB Higgs, we need to extend the model in some way. 
In the composite framework, one invokes some further explicit breaking of the global symmetry induced by e.g. the underlying fermion-mass mechanism. With elementary scalars, there is also another possibility, as described in~\cite{Alanne:2016mmn}: We can obtain a non-trivial vacuum alignment by adding a singlet scalar acquiring some of its mass from the vev of $\Phi$. The most minimal extension of this kind is to add a real $Z_2$-symmetric singlet scalar and respectively extend the scalar potential to
\begin{equation}
    \label{eq:}
	V_0=\frac{1}{2}m^2 \Phi^{\dagger}\Phi+\frac{1}{2}m_S^2S^2+\frac{\lambda}{4!}(\Phi^\dagger \Phi)^2
	    +\frac{\lambda_{\sigma S}}{4}(\Phi^\dagger\Phi)S^2+\frac{\lambda_S}{4!}S^4.
\end{equation}

In the limit of very heavy singlet, $M_S\ll v$, we find 
\begin{equation} 
\label{eq:lambda-theta}
	\sin^2\theta =  \lambda_{\sigma S}\frac{v_{w}^2  \left( 3 A + 2 B + 2 A \log \dfrac{g^2 v_{w}^2}{M_S^2}\right)}{4 M_S^2 \left( 2 A + B + A \log  \dfrac{g^2 v_{w}^2}{M_S^2}\right) } + {\cal O} \left(\left({v/M_S}\right)^{4}\right),
\end{equation}
where
\begin{align}
    \label{eq:}
    A=\frac{1}{16g^4}&\left(3g^4+2g^2g^{\prime\,2}+g^{\prime\,4}-16y_t^4\right),\nonumber\\
    B=\frac{1}{16g^4}&\left[(g^2+g^{\prime\,2})^2\log\left(\frac{1}{4}\left(\frac{g^{\prime\,2}}{g^2}+1\right)\right)-4g^4\log2-16y_t^4\log\frac{y_t^2}{2g^2}\right.\\
    &\left.-\frac{5}{6}\left(3g^4+2g^2g^{\prime\,2}+g^{\prime\,4}\right)+24y_t^4\right].\nonumber
\end{align}
It is worth noting that this reproduces the EW symmetry in the excactly decoupled limit. For finite values of $M_S$, the portal coupling is directly responsible for a non-zero value of $\theta$. 

Another possibility is to add an additional global-symmetry-breaking term in the similar manner as in the composite framework. The minimal of this kind is given by
\begin{equation}
	V_B = C_B v^3 E_0^\dagger \Phi,
\end{equation}
where $C_B$ is a dimensionless constant.

    \subsection{Comparison with the composite case}
    \label{sec:comp}

The leading EW-gauge and fermion-Yukawa contributions to the effective potential in the composite framework 
are given by:
\begin{equation}
    \label{eq:}
\widetilde{V}^{\text{gauge}}_1= \frac{3}{64\pi^2}\sum_K C_K\mathrm{Tr}\left[ \mu^2\right]\Lambda^2,\quad \text{ and } \quad
\widetilde{V}^{\text{ferm}}_1= -\frac{4}{64\pi^2}\sum_F C_{F}\mathrm{Tr}\left[ m_F^{\dagger}m_F\right]\Lambda^2,
\end{equation}
where $\Lambda$ is the physical cut-off, and $C_K,C_F$ are the form factors related to gauge group $K$ and fermion $F$, resp.
Identifying $\Lambda$ with the compositeness scale $\Lambda\sim 4\pi f$, where  $f$ is the ``pion decay constant'' 
for the underlying composite model, we obtain for the SM gauge group and top-quark embedding:
    \begin{align}
	\label{eq:VeffComp}
	    \left.\widetilde{V}^{\mathrm{gauge}}_1\right|_{\mathrm{vac}} &= \frac{3}{16}\left(3 g^2 C_g +g^{\prime 2}C_{g^{\prime}}\right)\sin^2\theta f^4,\\
	    \left.\widetilde{V}^{\mathrm{top}}_1\right|_{\mathrm{vac}}&= -\frac{3}{2}y_t^2 C_t\sin^2\theta f^4.
    \end{align} 
    The gauge part has a minimum at $\theta=0$, and thereby the gauge sector prefers to be unbroken as already noticed by  
		Peskin~\cite{Peskin:1980gc} and Preskill~\cite{Preskill:1980mz}. 
The top sector, on the other hand, prefers the minimum to be at $\theta=\pi/2$. 
This contribution was considered only later in recent technicolour and composite-Higgs models, see e.g. \cite{Galloway:2010bp,Cacciapaglia:2014uja}. Since the top contribution dominates over the gauge part, it aligns the vacuum in the direction where the electroweak symmetry is fully broken. Therefore, to achieve a pGB Higgs scenario, new sources of vacuum misalignment are needed. 

A minimal solution to achieve the desired vacuum alignment is to add an explicit symmetry breaking operator taking the form:
\begin{equation}
	V_B = - C_B f^4 E_0^\dagger \Sigma=- C_B f^4 \cos\theta+\dots ,
\end{equation}
where $C_B$ is a positive dimensionless  constant.

\section{Unification with radiative Fermi scale}
\label{sec:unification}

Unification of (some of) the SM interactions is one of the long-standing paradigms beyond the SM. These scenarios, however, imply a large  hierarchy between the EW symmetry breaking at the Fermi scale, $\vw=246$ GeV, and the breaking of the unified symmetry at some much higher scale. These symmetry breakings are typically modelled via ad-hoc scalar sectors, and there is no symmetry reason to prohibit the portal interactions between these two sectors. However, unless the portal coupling is highly suppressed compared to the SM Higgs self-coupling, the symmetry breaking at the unification scale would induce a large mass for the SM Higgs already at the tree level.

In~\cite{Alanne:2015fqh} we proposed a framework to explain this large hierarchy by vacuum  misalignment. The minimal scenario utilises global-symmetry-breaking pattern $\SU(4)\cong\SO(6)\rightarrow \Sp(4)\cong\SO(5)$ in the Pati--Salam-unification~\cite{Pati:1974yy}  framework. This scheme  unifies quarks and leptons by promoting the lepton number to the fourth colour, and therefore the full symmetry group is then $\SU(4)_{\mathrm{PS}}\times\SU(4)_{\mathrm{global}}$. In this scenario, the proton does not decay via gauge interactions, as is the case in the  Georgi--Glashow~\cite{Georgi:1974sy} framework, where one unifies strong and electroweak interactions. Instead,  the spin-one leptoquarks mediate $K_{\mathrm{L}}\rightarrow\mu^{\pm} e^{\mp}$. Strong experimental limits on these decays lead to a lower bound $M>1.5\cdot 10^6$~GeV on the leptoquark masses~\cite{Parida:2014dba} translating into the lower bound on the Pati--Salam-unification scale $\Lambda_{\mathrm{PS}}\gtrsim 1.9\cdot 10^6$~GeV.

	The $\SU(4)/\Sp(4)$ model with elementary scalars has been studied in both non-supersymmetric~\cite{Alanne:2014kea,Gertov:2015xma} and supersymmetric frameworks~\cite{Alanne:2016uta}. The breaking pattern is obtained by a scalar, $M$, transforming in the six-dimensional antisymmetric representation of $\SU(4)$, and can be conveniently parameterised in terms of the GBs, $\Pi^a$, and the $\Sp(4)$ singlet, $\sigma$, as
	\begin{equation}
	    \label{eq:}
	    M=(\frac{\sigma}{2}+\ii\sqrt{2}\Pi^aX^a)E,
	\end{equation}
	where $X^a$ are the broken generators with respect to vacuum $E$.
	
	We embed $\SUL\times\SUR\cong\SO(4)$ into $\SU(4)$ by identifying the left and right generators
	    \begin{equation}
		\label{eq:gensCust}
		T^i_{\mathrm{L}}=\frac{1}{2}\left(\begin{array}{cc}\sigma_i & 0 \\ 0 & 0\end{array}\right),\quad\text{and}\quad
		T^i_{\mathrm{R}}=\frac{1}{2}\left(\begin{array}{cc} 0 & 0 \\ 0 & -\sigma_i^{T}\end{array}\right),
	    \end{equation}
	    where $\sigma_i$ are the Pauli matrices. The generator of the hypercharge is then identified with the third generator 
	    of the $\SU(2)_{\mathrm{R}}$ group, $T_Y=T^3_{\mathrm{R}}$. After embedding the EW group, we can identify the EW preserving and breaking vacua, $E_0$ and $\EB$, resp.:
    \begin{equation}
	E_{0}= \left(\begin{array}{cc} \ii \, \sigma_2 & 0\\
	0 & -\ii \, \sigma_2 \\
	\end{array}\right), \quad \EB=\left(\begin{array}{cc} 
	0 & 1\\
	-1& 0\\
	\end{array}\right).
    \end{equation}
As discussed in Sec.~\ref{sec:SON} , the misalignment between the EW group and the stability group, $\Sp(4)$, can be parameterised by and angle $\theta$, and the actual vacuum can again be written as $\EE=\cos\theta E_0+\sin\theta\EB$.

To break the $\SU(4)_{\mathrm{PS}}$-leptocolour group to $\SU(3)_{\mathrm{c}}\times U(1)_{\mathrm{B}-\mathrm{L}}$, we need to introduce an addtional scalar multiplet. The minimal extension of the scalar sector contains a scalar multiplet, $P=p_aT^a$, transforming in the adjoint representation under the leptocolour group but is singlet under the global $\SU(4)_{\mathrm{glo}}$. This additional scalar multiplet thus, in addition to breaking the leptocolour, serves as the minimal EW-singlet extension allowing for a non-tirivial vacuum alignment as discussed in Sec.~\ref{sec:singl}. The most general renormalisable scalar potential including these scalar multiplets then reads
    \begin{equation}
	\label{eq:pot}
	\begin{split}
	    V=&\frac{1}{2}m_M^2\mathrm{Tr}[M^{\dagger} M]+m_P^2\mathrm{Tr}[P^{2}]+\frac{\lambda_M}{4}\mathrm{Tr}[M^{\dagger}M]^2
		+\lambda_{P1}\mathrm{Tr}[P^2]^2\\
	    &+\lambda_{P2}\mathrm{Tr}[P^4]+\frac{\lambda_{MP}}{2}\mathrm{Tr}[M^{\dagger}M]\mathrm{Tr}[P^2],
	\end{split}
    \end{equation}
and the desired breaking pattern occurs as the scalars acquire vevs $\langle M \rangle=\frac{v_0}{2}E$, and $\langle P \rangle=b_0 T^{15}$.

To find the actual vacuum alignment angle, we calculate the one-loop Coleman--Weinberg potential and minimize the full effective potential. We fix the renormalization scale such that the tadpole contributions in the $\sigma$ direction vanish, i.e. the vev $v=\langle\sigma\rangle$ is given by the tree-level value $v_0$, while the other vev, $\langle p_{15}\rangle=b$, is determined by minimising the full one-loop potential along with the dynamical value of $\theta$.

Fixing the mass of the Higgs boson further constrains the parameter space. Three states, $\Pi_4$, $ \sigma$ and $p_{15}$, have the same quantum numbers as Higgs, and the lightest eigenstate, turning out to be dominantly the pGB $\Pi_4$, is identified with the Higgs boson. 
The Higgs phenomenology constraints were investigated previously~\cite{Alanne:2014kea,Gertov:2015xma}, and it was shown that the elementary-Goldstone-Higgs paradigm reproduces the phenomenological success of the SM. Similar analysis applies here.

The numerical analysis shows that a  small value of $\theta$ is preferred, implying a mostly pGB Higgs boson. Furthermore, we find that the preferred values of $v$ are roughly of the order of $b$, which was fixed just above the experimental bound for the unification scale,  $b=2.5\cdot 10^6$~GeV. We checked that this conclusion is independent of the specific value of $b$, and the same conclusion holds for larger values of the unification scale as well.
Furthermore, the values of the quartic couplings are overall very small, and in particular there is no large hierarchy between them.
This feature of tiny quartic couplings originates from the relation between the couplings of the scalar potential and the vacuum angle $\theta$ given by the minimisation conditions. In the limit of equal self-couplings, and $v=b$, this relation is given by $\lambda\propto\sin^2\theta$. Note that this coincides with Eq.~\eqref{eq:lambda-theta}.

\section{Conclusions}
\label{sec:conc}

We argued that the different UV structures of a composite-Higgs scenario and a renormalisable model with elementary scalars lead to different vacuum alignment phenomenology in the presence of enhanced global symmetries. Furthermore, in the elementary-scalar case, it is possible to achieve a pGB-like Higgs by extending the scalar sector with an EW-singlet without introducing further explicit global-symmetry-breaking operators.

This provides an interesting connection between the Fermi scale physics and unification, since the scalar multiplet that breaks the unification symmetry generally couples to the EW-breaking scalar multiplet via portal interactions. Therefore, the heavier scalar multiplet can additionally play an important role in aligning the vacuum towards the pGB-like Higgs. We have provided a concrete model example in the Pati--Salam-unification framework and found out that once the quantum effects are taken into account, the enhanced global symmetry of the Higgs sector indeed tends to break near the unification scale, and the actual Fermi scale is radiatively generated via the vacuum-misalignment phenomenon. Another interesting feature of this kind of scenario is that all the quartic scalar couplings are generally tiny, but there is no large hierarchy between them. This opens interesting paths to be explored in the future ranging from neutrino masses to cosmic inflation.

\section*{Acknowledgments}
I thank the organizers of the 5th International Conference on New Frontiers in Physics for the opportunity to present this work. The CP$^3$-Origins centre is partially funded by the Danish National Research Foundation, grant number DNRF90. The author acknowledges partial funding from a Villum foundation grant.

\end{document}